\documentstyle[12pt]{article}
\begin{document}
\setlength{\baselineskip}{22pt}
\begin{titlepage}
\vspace{35mm}

\begin{center}
{\Large\bf {Stable distributions in fragmentation processes}}
\end{center}
\vspace{5mm}
\begin{center}
{G J Rodgers and M K Hassan$^*$}
\end{center}
Department of Physics, Brunel University, Uxbridge, Middlesex, UB8  3PH, 
United Kingdom

\vspace{30mm}

\begin{center}
Abstract
\end{center}

\noindent
We introduce three models of fragmentation in which the largest fragment 
in the system can be broken at each time step with a fixed probability, 
$p$. We solve these models exactly in the long time limit to reveal 
stable time invariant (scaling) solutions which depend on $p$ and the 
precise details of the fragmentation process. Various features of these 
models are compared with those of conventional fragmentation models

\vspace{20mm}

\noindent 
PACS: 0520.-y,0250.-r

\vspace{4mm}

\vspace{4mm}

\noindent
$*$  \hspace{4mm} \footnotesize {Permanent address: Department of Physics, 
Shahjalal Science and Technology University, Sylhet, Bangladesh}
\end{titlepage}
 
\section{Introduction}

The generation of random fragments by successive and sequential breaking 
is an irreversible kinetic process that occurs in a wide range of 
physical systems in both science and technology [1-6]. This variety of 
applications has motivated theorists to obtain analytical solutions to 
simple fragmentation models as a function of both the fragmentation rules 
and the initial conditions [7-10].

In this letter we introduce three unusual models of fragmentation which, 
at each time step, allow the fragmentation of the largest fragment in the 
system with some externally tuneable probability. In Model A the largest 
fragment is broken into two pieces at every time step. The size of the 
resultant fragments is determined by an arbitrary probability 
distribution. In Model B, the largest fragment is split with probability 
$p$ and with probability $(1-p)$ another fragment is chosen. When the 
later choice is made, every fragment is equally like to be chosen. In 
Model C, this choice is controlled by a homogeneity index $\beta$. These 
models are solved exactly in the long time limit to reveal stable time 
invariant probability distributions. The results are compared with 
conventional models of fragmentation.

This work was originally motivated by three different observations. 
Firstly, the extremal properties of fragmenting systems have been the 
subject of interest recently [11,12]. This work considered systems in 
which a fragment is put to one side at each time step and takes no 
further part in the fragmentation processes. This process leads to 
singularities in the probability distribution for the largest fragment 
[11]. 

Secondly, the kinetics of systems of growing and coalescing (equal-sized) 
droplets has been considered in which the 
two droplets which were closest together coalesced at the next time step 
[13]. This lead to some interesting kinetic behaviour, in particular the 
droplet patterns that emerged were self-similar in time with 
inter-droplet distances obeying simple scaling behaviour. We wondered, if 
the analogous system in fragmentation models would yield similarly 
interesting results.

Finally, we observed that the largest particle was always the most likely 
to fragment in any non-shattering fragmentation process. If we allowed 
the largest fragment to be chosen for fragmentation with some fixed 
probability this would allow us to utilise the methods in [13] to obtain 
solutions to the kinetic process. This fixed probability may seem a 
little unrealistic, but it is not difficult to envisage a kinetic system, 
such as froth under some external pressure, in which the largest bubble, 
with its greater volume  and surface area, is much more likely to 
fragment than any of the other bubbles in the system.

In the remainder of this introduction we give a brief summary of previous 
work on models of binary fragmentation. Then in the next three sections 
we introduce the models and present our analytical solutions. In section 
5, we discuss the scaling behaviour of the models and in the final 
section summarise our findings  and conclusions.

In most of these models of binary fragmentation, the concentration of 
fragments of size $y$ at time $t$, $n(y,t)$, evolves according to 
\begin{equation}
n(y,t+\delta t)=n(y,t)-\delta t n(y,t)\int_0^y R(z,y-z)dz +2\delta 
t)\int_y^{L(t)} R(y,z-y)n(z,t)dz
\end{equation}
where, $R(y,z)$ is the intrinsic rate at which a particle of size $(y+z)$ 
breaks into particles of size $y$ and $z$. The first term on the right 
hand side is the contribution from those particles not chosen for 
fragmentation in time $(t,t+\delta t)$. The second term represents the 
decrease in the number of particles of size  $y$ by fragmentation into 
particles of size $z$ and $(y-z) (y>z)$. The third term represents the 
increase in the number of particles of size $y$ due to the fragmentation 
of particles of size $z (>y)$, such that one of the products is of size 
$y$. 

The upper limit of the second integral on the right hand side is usually 
set to some fixed value greater than the size of the largest particle in 
the system. This value is usually $1$ or $\infty$ depending on the 
details of the model. To connect with the analysis of our models  in the 
following sections, we have set the upper limit to $L(t)$, the size of 
the largest particle at time $t$. As $n(y,t)=0$ for all $y>L(t)$, this 
choice has no effect on our results.

A general scaling theory has been constructed for systems of 
fragmentation particles [10]; the particle size distribution in the limit 
$y\rightarrow 0, t\rightarrow \infty$ can be written as 
\begin{equation}
n(y,t) \sim s(t)^{-2}\Phi({{y}\over{s(t)}})
\end{equation}
where $s(t)$ is the time dependent typical cluster mass and the exponent 
$2$ is required by mass conservation. Exact solutions have been obtained 
to the kinetic equation (1) with a number of different fragmentation 
rules $R(y,z)$. In particular, solutions have been obtained [3,8,9] when 
$R(y,z)=(y+z)^{\omega -1}$. This describes a situation in which 
the rate of fragmentation is determined solely by the size of 
the particle and every point along the line of the particle is 
equally likely to fragment. When $\omega>0$ this system 
exhibits simple scaling dynamic scaling and $s(t) \sim t^{-z}$ 
with $z=1/\omega$. For $\omega<0$ shattering occurs and a 
cascading break-up of particles causes a finite amount of mass 
to be transferred to an infinite number of particles with zero 
mass [8,10].
We now turn to detailed consideration of our models.

\section{Model A}

Firstly, we consider a model in which the largest particle is fragmented 
into two pieces at each time step. If we assume that in time $(t,t+\delta 
t)$ the size of the largest fragment changes from $L(t)$ to 
$(L(t) -\delta L)$ then we can choose $R(y,z)$ as 
\begin{equation}
\delta tR(y,z) = {{\delta L}\over 
{L(t)}}f({{y}\over{L(t)}})\delta(y+z-L(t))
\end{equation}
This term is easily understood; the $\delta$-function ensures 
that only the largest particle is fragmented, $f(x)$ is defined 
as the rate with which particles of size $xL$ and $(1-x)L$ are 
created from the largest particle $L(t)$ and ${{\delta 
L}\over{L(t)}}$ is the probability of placing the cut in the 
largest particle in a particular infinitesimal length $\delta 
L$. This rate is time dependent, through its dependence on 
$L(t)$, which is in line with the definition of the model. Inserting the 
rate into the kinetic equation (1) yields
\begin{equation}
n(y,t+\delta t)=n(y,t)+{{2\delta L}\over{L(t)}}n(L(t),t)f({{y}\over{L(t)}})
\end{equation}
where the second term on the right-hand side describes the gain in the 
number of particles of size $y$ from the fragmentation of the 
largest fragment.

We can define the normalised density of fragments of length $y$ 
at time $t$ as 
\begin{equation}
g(y,t)={{n(y,t)}\over{\int_0^{L(t)}n(z,t)dz}}
\end{equation}
so that, as $L(t)\rightarrow L(t)-\delta L$ in time $t 
\rightarrow t+\delta t$,
\begin{equation}
g(y,t+\delta t)={{n(y,t\delta 
t)}\over{\int_0^{L(t)-\delta L}n(z,t+\delta t)dz}}.
\end{equation}
Consequently, following the methods of [13], we can rewrite equation (4) 
as 
\begin{equation}
g(y,t+\delta t)=g(y,t)[1-\delta L g(L,t)]+{{2\delta 
L}\over{L}}g(L,t)f({{y}\over{L}})
\end{equation}
The natural length scale in this model is $L(t)$, the length of 
the largest fragment. So, to solve equation (7), we take 
account of this by introducing $F(x,t)$, defined by,
\begin{equation}
F(x,t)=L(t)g(xL(t),t)
\end{equation}
Notice that $x$, defined by $x=y/L(t)$, is restricted to the 
range $[0,1]$. Remembering that $L(t)$ is a function of time , 
we can obtain a partial differential equation for $F(x,t)$;
\begin{equation}
{{\partial F(x,t)}\over{\partial 
t}}=2F(1,t)f(x)-F(1,t)F(x,t)-F(x,t)-x{{\partial F(x,t)}\over{\partial x}} 
\end{equation}
Here, we have set $\delta x/\delta t$, which determines the relationship 
between the real time and the length of the largest fragment $L(t)$, to 
$1$. This choice is arbitrary and has no effect on the kinetics of the 
process.

The remainder of this section is concerned with solutions to this 
equation. First of all we will assume that $F(x,t)$ evolves to a time 
independent quantity as $t\rightarrow \infty$. In this limit we can solve 
(9) to yield
\begin{equation}
F(x)={{F(1)}\over{x^{F(1)+1}}}-2{{F(1)}\over{x^{F(1)+1}}}\int_x^1f(y)y^{F(1)} dy
\end{equation} 
where $F(x)=\lim_{t\rightarrow \infty}F(x,t)$.

There are a number of constraints on $f(x)$. Obviously to 
preserve symmetry we must have $f(x)=f(1-x)$ and the 
distribution is normalised so that
\begin{equation}
\int_0^1f(y)dy=1.
\end{equation}
These two constraints lead to 
\begin{equation}
\int_0^1yf(y)dy=1/2.
\end{equation}
A choice of $f(x)$ which satisfies both these constraints is 
\begin{equation}
f(x)={{x^\alpha(1-x)^\alpha}\over{\int_0^1y^\alpha(1-y)^\alpha dy}}
\end{equation}
where we assume $\alpha >-3/2$ so that the variance of $f(x)$ is well 
defined. Inserting (13) into (10) and integrating $F(x)$ 
between $0$ and $1$ reveals
\begin{equation}
\int_0^1F(x)dx=1+[1-2{{\beta(\alpha+F(1)+1,\alpha+1)}\over{\beta(\alpha+1,
\alpha+1)}}]\lim_{x\rightarrow 0} x^{-F(1)}
\end{equation}
For $F(x)$ to be correctly normalised the second term on the 
right hand side must be zero. The condition $F(1)>1$ means that 
this term diverges unless the contents of the square bracket 
are zero. It is a simple matter to show that this can be 
achieved for all $\alpha$ by setting $F(1)=1$. As a result, we 
rewrite equation (10) as 
\begin{equation}
F(x)={{1}\over{x^2}}[1-2\int_x^1yf(y)dy].
\end{equation}
Substituting (13) into equation (15) and evaluating the 
integral reveals that for $\alpha=0$ we have the trivial 
solution $F(x)=1$ for all $x$. For $\alpha=1$ we have 
\begin{equation}
F(x)=x(4-3x).
\end{equation}
for $\alpha=2$,
\begin{equation}
F(x)=x^2(15-24x+10x^2)
\end{equation}
and for $\alpha=-1/2$
\begin{equation}
F(x)={{1}\over{x^2}}-{{2}\over{\pi x^2}}[Cos^{-1}(\sqrt{x})+\sqrt{x(1-x)}].
\end{equation}
We checked these solutions numerically by calculating 
$F(x,t_0)$ at some (long) time $t_0$ for a number of 
realisations of the randomness in the system and then averaging 
$F(x,t_0)$ over a large number of these realisations. The 
average values of $F(x,t_0)$ we obtained agreed with the 
analytical solutions in equations (15)-(18). 

We can also consider the behaviour as $x\rightarrow 0$. If we 
assume that $x\rightarrow 0, f(x) \rightarrow Ax^\beta$ then we 
find that
\begin{equation}
F(x) \rightarrow {{2A}\over{\beta + 2}}x^\beta
\end{equation}
in the same limit. If we take $f(x)$ to have the form (13) then 
we have $\beta=\alpha$ and 
$A=\Gamma(2\alpha+2)/(\Gamma(\alpha))^2$. Alos, notice that when 
$\alpha<0$, $F(x)$ diverges as $x\rightarrow 0$ and when 
$\alpha > 0$, $F(x)$ goes to zero smoothly. This behaviour in 
the probability distribution at $x=0$ is similar to that 
observed in other models [3,10].

The above analysis is restricted to the stationary state. 
However, as these models are relatively simple, one can attempt 
to obtain explicit time dependent solutions. For instance, 
using the method of characteristics, one can solve equation (9) 
when $f(x)=1$ self-consistently to reveal
\begin{equation}
F(x,t)=2\int_0^tF(1,u)e^{u-t+\int_t^uF(1,v)dv}du+F(xe^{-t},0)e^{-t+\int_0^tF(1,v)dv}.
\end{equation}
One can then in principle, insert some initial conditions $F(x,0)$, to 
solve for $f(1,t)$ then substitute $F(x,0)$ and $F(1,t)$ into 
(20) to give the explicit solution. The only case we have been 
able to do this when $F(x,0)=1$ for $0 \leq x \leq 1$. This 
yields the trivial stationary solution $F(x,t)=1$. We were not 
able to solve (9) by the method of characteristics for any more 
complicated $f(x)$.

\section{Model B}

In this section we consider a model in which, at each time 
step, either the largest fragment is selected with probability 
$p$ or another fragment is chosen with probability $(1-p)$.

If the largest fragment is chosen, it is split into pieces $xL$ 
and $(1-x)L$ with a probability independent of $x$. That is to 
say the daughter distribution function is uniform, or in the 
formalism of Model A, $f(x)=1$ or $\alpha=0$. If it is decided 
to split a smaller fragment, every fragment is equally likely 
to be chosen for fragmentation. This corresponds, to the onset 
of shattering in conventional fragmentation models when 
$\omega=0$. Consequently, when $p=1$, this model is equivalent 
to Model A, when $p=0$, it is equivalent to a model of 
fragmentation originally solved by Ziff and McGrady [9].

The fragmentation rate for this model is given by
\begin{equation}
\delta tR(y,z)=(1-p){{\delta t}\over{y+z}}+p{{\delta 
L}\over{L(t)}}\delta(y+z-L(t))
\end{equation}
and the kinetic equation for the process is 
\begin{equation}
n(y,t+\delta t)=n(y,t)+2p{{\delta L}\over{L(t)}}n(L(t),t)-(1-p)\delta 
tn(y,t)+2(1-p)\delta t\int_y^{L(t)}n(z,t){{dz}\over{z}}.
\end{equation}
Using an identical method to that used for Model A we can obtain
\begin{equation}
{{\partial F(x,t)}\over{\partial t}} 
=2F(1,t)-[3-2p+F(1,t)]F(x,t)-x{{\partial F(x,t)}\over{\partial 
x}}+2(1-p)\int_x^1F(y,t){{dy}\over{y}}
\end{equation}
where we have set $p\delta x/\delta t=1$ and assumed that $p>0$. In the 
limit $t\rightarrow \infty$ we can solve this equation self-consistently 
for $p>0$ to reveal the time invariant solution
\begin{equation}
F(x) =px^{p-1}
\end{equation}
Here, in contrast to Model A, $F(1)$ is picked out by the solution of the 
equation and there is no need to consider the normalisation of $F(x)$ to 
find $F(1)$. As before, this result has been confirmed by numerical 
simulation. Models A and B coincide with the same special, trivial 
solution when $\alpha =0$ and $p=1$.

\section{Model C}

In this section we consider a generalised version of Model B in which the 
rate with which the non-largest fragments are chosen for fragmentation is 
controlled by a homogeneity index $\beta$. More precisely, a particle of 
size $y+z$ is split into fragments of size $y$ and $z$ with a rate 
$(y+z)^{\beta-1}$ [9]. This equivalent to $\omega=\beta$ in conventional 
models. Consequently, $R(y,z)$ is given by 
\begin{equation}
\delta tR(y,z)=(1-p)\delta t(y+z)^{\beta-1}+p{{\delta 
L}\over{L(t)}}\delta(y+z-L(t))
\end{equation}
and the rate equation for this process is 
\begin{equation}
n(y,t+\delta t)=n(y,t)+2p{{\delta L}\over{L(t)}}n(L(t),t))-(1-p)\delta
ty^\beta n(y,t)+2(1-p)\delta t\int_y^{L(t)}n(z,t)z^{\beta-1}dz
\end{equation}
In this model, if one sets $\beta=0$, we recover Model B and when $p=1$, 
we get Model A. 

We now follow the same steps as before to reveal an integro-differential 
equation for $F(x,t)$;
\begin{eqnarray}
{{\partial F(x,t)}\over{\partial t}} 
& = & 2\gamma F(1,t)-[\gamma (1+F(1,t))+(1-p)(x^\beta+\int_0^1F(y,t)y^\beta 
dy)]F(x,t) \nonumber \\  & - & \gamma x{{\partial F(x,t)}\over{\partial
x}} +2(1-p)\int_x^1F(y,t)y^{\beta+1}dy
\end{eqnarray}
where the time dependent function $\gamma(t)$ is defined by 
\begin{equation}
p{{\delta x}\over{\delta t}}=L(t)^\beta \gamma(t).
\end{equation}
We choose the relationship between the real time $t$ and the largest 
fragment $L(t)$ so that in the limit $t\rightarrow \infty$, $\gamma(t)$ 
goes to some constant, to be determined, $\gamma$. Consequently, in the 
long time limit when the partial time derivative of $F(x,t)$ goes to 
zero, equation (27) becomes
\begin{equation}
\gamma x{{\partial F(x)}\over{\partial x}} +[\gamma 
(1+F(1))+(1-p)(x^\beta+\int_0^1F(y)y^\beta dy)]F(x)=2\gamma 
F(1)+2(1-p)\int_x^1F(y)y^{\beta+1}dy
\end{equation}
This can be solved self-consistently using the functional form for $F(x)$;
\begin{equation}
F(x)=ae^{-bx^\beta}.
\end{equation}
Substituting (30) into (29) reveals two equations for $a$ and $b$,
\begin{equation}
a={{\beta 
b^{{{1}\over{\beta}}}}\over{\int_0^be^{-t}t^{{{1}\over{\beta}}-1}dt}}
\end{equation}
and 
\begin{equation}
b={{1-p}\over{\beta \gamma}}
\end{equation}
where the integral in the denominator of $a$ is an incomplete gamma 
function. As in Model B, we find that $F(1)=p$ is picked out as 
a condition for a function of the form (30) to be a solution. 
Equations (31) and (32), together with $F(1)=p$, completely 
determine $a, b$ and $\gamma$. So, as an example, we consider $\beta =1$, 
when we obtain an exponential $F(x)$
\begin{equation}
F(x)=pe^{b(1-x)}
\end{equation}
where
\begin{equation}
b=p(e^b-1).
\end{equation}
We have confirmed this solution, and those for some other values of 
$\beta$, by numerical simulation.

\section{Scaling}

In this section we consider the scaling properties of all three models. 
The function $F(x,t)$ is related to the number density of the fragments, 
$n(y,t)$, via 
\begin{equation}
n(y,t)={{N(t)}\over{L(t)}}F({{y}\over{L(t)}},t)
\end{equation}
where, $N(t)$ is the total number of fragments. Notice that the total 
length of the fragments, a constant, is given by
\begin{equation}
\int_0^{L(t)}yn(y,t)dy=N(t)L(t)\int_0^1xF(x,t)dx.
\end{equation}
In the limit $t\rightarrow \infty$, when the integral on the right hand 
side becomes time independent, we know that $N(t) \sim t$, so we can 
deduce that $L(t) \sim 1/t$. Inserting this result into equation (35) 
gives the scaling form,
\begin{equation}
n(y,t) \sim t^2F({{y}\over{L(t)}}) \sim t^2 \Phi(yt)
\end{equation}
indicating that the typical cluster mass $s(t) \sim 1/t$ and the kinetic 
exponent $z=1$ for all these models. So, in the scaling limit 
$t\rightarrow \infty$, the scaling function $\Phi(\xi)\sim F(\xi)$, where 
$\xi \sim yt$.

To illustrate this scaling result, and to show the agreement with our 
analytical result, we have done two numerical simulations. In figure 1, 
we have shown  a plot of $n(y,t)/t^2$ against $yt$ from a simulation of 
Model B with $p=1/3$ for three different times. This figure can be 
thought of as a plot of $\Phi(\xi)$ against $\xi$ once each axis has been 
rescaled by constant prefactors. We see that the data collapses as 
predicted and in line with our analytical results (equation (19)), 
$\Phi(\xi) \sim \xi^{-2/3}$ for $\xi<1$. The analytical result is shown 
as solid line on the figure. The region $\xi >1$, where $\Phi$ is zero 
and the point $\xi=1$, where $\Phi$ goes to zero discontinuously, is not 
shown.

In figure 2 there is similar plot for Model A with $\alpha =1$. After 
rescaling both axes  by suitable prefactors this result exactly matches 
the analytic prediction in equation (13); $\Phi(\xi) \sim \xi(4-3\xi)$ 
for $\xi<1$ and $\Phi(\xi) =0$ for $\xi>1$. This prediction is shown with 
the solid line. In this figure we have included the region around 
$y\approx  L(t) \approx 1/t$ (about $yt =0.034$ in the units on the figure, 
or $\xi=1)$ where the data collapse is not quite so good. This is due to 
the difficulty of capturing the discontinuity at $y\approx L(t) (\xi =1)$ 
numerically. However, one can see that the slope of the curve around 
$y\approx L(t)$ gets progressively steeper as $t$ gets larger. This is as 
expected; the discontinuity would be captured exactly in the limit 
$t\rightarrow \infty$.

In the above discussion we saw that the largest fragment $L(t) \sim 
t^{-\delta}$ with $\delta =1$ in the long time limit. To compare this 
particular result with conventional models of fragmentation in which a 
particle of size $y+z$ is split into fragments of size $y$ and $z$ with a 
rate equal to $1/(y+z)(\omega=0)$ or independent of $y$ and $z 
(\omega=1)$. In these two models, we found that the size of the largest 
fragments also decays algebraically, but more slowly, as $t{-\delta}$ 
with $\delta =0.20$ and $0.85$ for $\omega=0$ and $1$ respectively.

\section{Summary and Conclusions}
 
To briefly summarise our results, we have introduced three models of 
fragmentation in which there is a fixed probability per time step of 
breaking the largest fragment in the system. In Model A the largest 
fragment was fragmented at each time step with an arbitrary fragmentation 
kernel. In Model B and C the largest fragment was chosen for 
fragmentation with probability $p$ and $(1-p)$ another fragment was 
chosen. The behaviour of these models is completely described by the 
function $F(x,t)$, which evolves to a stable distribution $F(x)$ in the 
limit $t\rightarrow \infty$. We were able to obtain an analytic solution 
for $F(x)$ for a number of different fragmentation kernels. We found that 
the models exhibited simple scaling , with kinetic exponent $z=1$  and 
the scaling function was equal to $F(x)$.

To study the kinetics of these models we looked for solutions of the 
kinetic equation obeyed by $F(x,t)$. In the simplest case of Model A 
(equation (9)), this is a non-linear in $t$, non-local in $x$, partial 
equation of the two variables $t$ and $x$. This means that solutions must 
be obtained self-consistently and this limits the scope for finding exact 
time dependent solutions to all but the most trivial problems. Our 
inability to study the kinetics analytically leads one to wonder about 
the range of validity of the stationary solutions we have obtained. Do 
all initial conditions lead to these stationary solutions, or if not, how 
is the phase space of initial conditions divided up? We have performed a 
number of numerical simulations and have always found that these models 
evolve to the predicted stationary states. We suspect that all well 
behaved, physically important, smooth initial conditions evolve to these 
stationary states. However, if we are able to find a set of initial 
conditions that do not evolve to these stationary states, we will return 
to this subject in a future publication.

\vspace{6mm}
  
\begin{flushleft}
{\large\bf Acknowledgement}

\vskip5mm

MKH would like to thank the CVCP for their financial support.

\newpage
\end{flushleft}
\begin{flushleft}
{\large\bf References}

\vskip5mm

1. Basedow A M, Ebert K H  and Ederer H J, 1978 
Macromolecules {\bf 11}, 774.

2. Ballauf M and Wolf B A 1981 Macromolecules {\bf 14}, 654.

3.  Ziff R M and McGrady E D, 1986 Macromolecules {\bf 19}, 2513.

4. Gilvarry J J 1961 J Appl. Phys. {\bf 32} 391.

5. Shinnar R 1961 J. Fluid Mech. {\bf 10} 269.

6.  Meyer R, Almin K E, and Steenberg B, 1966 Br. J. Appl. Phys. {\bf 17},
409.

7. Filippov A F, 1961 Theor. Prob. Appl. (USSR) {\bf 6}, 275.

8.  McGrady E D and Ziff R M, 1987 Phys. Rev. Lett. {\bf 58}, 
892.

9.  Ziff R M, McGrady E D  1990 J. Phys. A: Math. Gen. {\bf 18}, 3027

10.  Cheng Z and Redner S, 1990 J. Phys. A: Math. Gen. {\bf 23}, 1233.

11. Derrida B and Flyvbjerg H 1987 J. Phys. A: Math. Gen. {\bf 20} 5273.

12. Frachebourg L, Ispolatov I and Krapivsky P L 1995 Physical Rev. E 
{\bf 52} R5727.

13. Derrida B, Godreche C and Yekutieli I 1990 Europhys. Lett. {\bf 12} 
385.
\end{flushleft}

\newpage
\begin{flushleft}
Figure Captions

1. \hspace{5mm} $n(y,t)/t^2$ against $yt$ for Model B with $p=1/3$ for 
three different times. The solid line is a plot of the analytical result.

2.\hspace{5mm} As figure 1 for Model A with $\alpha =1$.
\end{flushleft}
   \end{document}